\documentclass[aps,
	       prl,
	       nofootinbib,
	       nobibnotes,notitlepage,
	       ]{revtex4-1} 

\usepackage{amssymb}
\usepackage{array}
\usepackage{multirow}
\usepackage{amsmath}
\usepackage{graphicx,subfigure}
\usepackage{longtable}
\usepackage{verbatim}
\usepackage{amsfonts}
\usepackage{graphicx}
\usepackage{hyperref}
\usepackage{xcolor}
\usepackage[normalem]{ulem}

\allowdisplaybreaks

\newcommand{\ov}{\overline}

\newcommand{\be}{\begin{equation}}
\newcommand{\ee}{\end{equation}}
\newcommand{\bea}{\begin{eqnarray}}
\newcommand{\eea}{\end{eqnarray}}
\newcommand{\beq}{\begin{equation}}
\newcommand{\eeq}{\end{equation}}
\def\beqa{\begin{eqnarray}}
  \def\eeqa{\end{eqnarray}}
\newcommand{\bv}{\left(\begin{array}{c}}
\newcommand{\ev}{\end{array}\right)}

\def\lsim{\mathrel{\rlap{\lower4pt\hbox{\hskip1pt$\sim$}}
    \raise1pt\hbox{$<$}}}	  
\def\gsim{\mathrel{\rlap{\lower4pt\hbox{\hskip1pt$\sim$}}
    \raise1pt\hbox{$>$}}}	  

\newcommand{\GeV}{\,\mbox{GeV}}

\begin{document}


\title{Data determination of HQET parameters in inclusive charm decays}
\author{Kang-Kang Shao$^{a,b}$}
\email{shaokk18@lzu.edu.cn}
\author{Chun Huang$^c$}
\email{chun.h@wustl.edu}
\author{Qin Qin$^a$}
\email{qqin@hust.edu.cn}
\affiliation{$^a$School of physics, Huazhong University of Science and Technology, Wuhan 430074}
\affiliation{$^b$Frontiers Science Center for Rare Isotopes, and School of Nuclear Science and Technology, Lanzhou University, Lanzhou 730000, China}
\affiliation{$^c$Physics Department and McDonnell Center for the Space Sciences, Washington University in St. Louis; MO, 63130, USA}
\vspace*{1cm}

\begin{abstract}
This work delves into the phenomenology of electronic inclusive decays of $D$ mesons, encompassing $D^0, D^+, D^+_s\to Xe^{+}\nu$. The theoretical formulas for the decay widths and electron energy moments of these decays are presented as expansions with powers of $\alpha_s$ and $\Lambda_{\rm QCD}/m_c$. Remarkably, the expansion exhibits excellent convergence properties when we choose the 1S mass scheme for charm. The formulas are subsequently fitted to experimental data, and the $D$ meson matrix elements of operators in the heavy quark effective theory are hence determined by data for the first time, including 
\begin{align}
\mu^2_\pi(D^{0,+}) &= (0.09\pm 0.05) \mathrm{GeV}^2, \qquad \qquad
\mu^2_\pi(D^{+}_s) = (0.11\pm 0.05) \mathrm{GeV}^2,  \nonumber \\
\mu^2_G(D^{0,+}) &= (0.32\pm 0.02) \mathrm{GeV}^2, \qquad \qquad 
\mu^2_G(D^{+}_s) = (0.43\pm 0.02) \mathrm{GeV}^2, \nonumber \\
\rho_D^3(D^{0,+}) &= (-0.003\pm 0.002) \mathrm{GeV}^3, \qquad\ 
\rho_D^3(D^{+}_s) = (-0.004\pm 0.002) \mathrm{GeV}^3, \nonumber \\
\rho_{LS}^3(D^{0,+}) &= (0.004\pm 0.002) \mathrm{GeV}^3, \qquad \ \ \ 
\rho_{LS}^3(D^{+}_s) = (0.005\pm 0.002) \mathrm{GeV}^3 . \nonumber 
\end{align}
These determined parameters will play a crucial role as inputs in various physical quantities, including $D$ meson lifetimes.
\end{abstract}

\maketitle
\section{Introduction}
Semi-leptonic inclusive heavy hadron decays present an optimal setting for precision testing of the standard model, owing to the systematic power expansion of the heavy quark mass within the framework of operator product expansion (OPE), which renders it theoretically more robust than exclusive decays. The inclusive charm decays such as $D\to X_{d,s}e^{+}\nu$, in addition to serving as complementary channels to inclusive beauty decays~\cite{Belle:2023asx,Belle-II:2023qyd,Huber:2019iqf,Huber:2020vup,Huber:2024rbw,Fael:2024rys,Fael:2024fkt}, hold intrinsic value. Despite criticisms regarding the slower convergence of the expansion in inclusive charm decays compared to beauty decays, an alternative perspective underscores a notable advantage: the enhanced impact of power corrections in charm decays allows for more efficient extraction of the corresponding nonperturbative parameters, which can be utilized in beauty decays through the principles of heavy quark symmetry.

The investigation of inclusive charm decays is further motivated by its close connection to several key issues of heavy flavor physics, including the determination of charmed hadron lifetimes, testing the Cabibbo-Kobayashi-Maskawa (CKM) mechanism, and addressing anomalies observed in $B$ meson decays. Notably, the calculation of charmed hadron lifetimes has yielded perplexing results, such as the observation of a negative central value for the $D^+$ decay width \cite{King:2021xqp} and the distinct ordering of charmed baryon lifetimes from experimental data \cite{Cheng:2021qpd,LHCb:2021vll,Belle-II:2023eii}. These discrepancies can be largely attributed to significant theoretical uncertainties resulting from imprecise and model-dependent input for non-perturbative hadronic parameters~\cite{Gratrex:2022xpm,Cheng:2023jpz}. In the realm of heavy quark effective theory (HQET), the theoretical framework governing the semi-leptonic inclusive decay rates and total decay widths of charmed hadrons relies on the same hadronic matrix elements of HQET operators. Referred to as HQET parameters, their precise determination from model-independent analyses of spectra from inclusive decays is crucial for resolving the lifetime anomalies observed in $D$ mesons and charmed baryons. Moreover, the longstanding discrepancy between the inclusive and exclusive determinations of $V_{cb,ub}$~\cite{HeavyFlavorAveragingGroupHFLAV:2024ctg} raises questions regarding whether this discrepancy arises from dynamics beyond the CKM mechanism or from overlooked QCD effects. A comparative analysis in the charm sector, specifically examining the inclusive and exclusive values of $V_{cd,cs}$, would be highly beneficial. This comparison hinges on the first determination of $V_{cd,cs}$ in semi-leptonic inclusive charm decays. Additionally, the intriguing $B$ anomaly observed in flavor-changing-neutral-current exclusive channels such as $B\to K^*\mu\mu$~\cite{Archilli:2017xmu} warrants exploration in the corresponding inclusive channels~\cite{Huber:2019iqf,Huber:2020vup,Huber:2024rbw}, including those in charm decays~\cite{deBoer:2015boa}. 

The CLEO experiment has conducted measurements of the electronic inclusive decays of the three types of $D$ mesons, 
$D^+,D^0,D_s^+\to e^+X$, providing data on branching ratios and electron energy spectra as reported in~\cite{CLEO:2009uah}. Recently, the BESIII collaboration updated the results for the $D_s^+$ channel ~\cite{BESIII:2021duu} and made the first and second measurements for $\Lambda_c\to e^+ X$~\cite{BESIII:2018mug, BESIII:2022cmg}. Comprehensive theoretical discussions on the heavy quark expansion for inclusive charm decays can be found in~\cite{Fael:2019umf}. Phenomenological studies include attempts of extracting weak annihilation contributions~\cite{Ligeti:2010vd, Gambino:2010jz} and determining the strong coupling constant $\alpha_s$~\cite{Wu:2024jyf} from experimental data. In addition, the potential for extracting the HQET parameters from current and future BESIII data has been explored in~\cite{Bernlochner:2024vhg}. By this work, we achieve for the first time extracting the HQET parameters by utilizing  available data. 

The remaining sections of this paper are structured as follows. In the subsequent section, we present the theoretical formulas for the decay widths and electron energy moments of the inclusive $D^+,D^0,D_s^+\to e^+X$ decays by expanding them in inverse powers of the charm mass using OPE. The leading power contributions, corresponding to the partonic results, are computed up to next-to-next-to-leading order (NNLO) of $\alpha_s$. We incorporate power corrections up to $1/m_c^3$, which are characterized by the HQET parameters, matrix elements of dimension-five and -six HQET operators, and their associated coefficients calculated to leading order (LO) of $\alpha_s$. In Section III, we extract the experimental results for the electron energy moments based on the corresponding spectra provided by CLEO and BESIII. Moving on to Section IV, we perform a data fitting of the theoretical formulas to the experimental data, determining the HQET parameters $\mu_\pi^2$, $\mu_G^2$, $\rho_D^3$, and $\rho_{LS}^3$, in inclusive charm decays. Finally, we conclude with a summary and prospect to possible theoretical and experimental improvements.

\section{Theoretical formulas}

In this section, we present the theoretical formulas for the inclusive $D\to e^+X$ decay widths and the electron energy moments, which are used for fitting the data. By employing OPE and perturbatively calculating the short-distance coefficients, these formulas are expanded in powers of $\Lambda_{\mathrm{QCD}} / m_c$ and $\alpha_s$, with the strange quark mass $m_s$ treated as a quantity of order $\Lambda_{\mathrm{QCD}}$ as in~\cite{Fael:2019umf}. We will consider power corrections up to $(\Lambda_{\mathrm{QCD}} / m_c)^3$. For the leading power contributions, we will include $\alpha_s$ corrections up to NNLO, while for power corrections, only LO $\alpha_s$ contributions will be considered. Various charm mass schemes, including pole mass, $\overline{\mathrm{MS}}$ mass, and 1S mass~\cite{Hoang:1998ng,Hoang:1998hm, Hoang:1999zc}, will be tested, and our analysis shows that the 1S mass scheme is the most favorable choice.



For the decay widths in the pole mass scheme, the results read 
\begin{align}\label{eq:decaywidth}
\Gamma_{D_{i}}=\sum_{q=d,s}\hat{\Gamma}_0\left|V_{cq}\right|^2 m_c^5 \Big\{ 1 &+\frac{\alpha_s}{\pi} {2\over3}\left(\frac{25}{4}- {\pi^2}\right) + \frac{\alpha_s^2}{\pi^2}\left[  {\beta_0\over 4} \frac{2}{3} \left(\frac{25}{4}-\pi ^2\right)\log \left(\frac{\mu^2}{m_c^2}\right) +2.14690 n_l-29.88311 \right]  \nonumber \\ 
  & -8\rho\delta_{sq} -\frac{1}{2}\frac{\mu_{\pi}^{2}(D_{i})}{m_{c}^2}-\frac{3}{2}\frac{\mu_{G}^{2}(D_{i})}{m_{c}^{2}} +6 \frac{\rho_D^3\left(D_{i}\right)}{m_c^3}+ ...\Big\},
\end{align}
where the constant $\hat{\Gamma}_0 = {G_F^2 }/{(192 \pi^3)}$, $V_{cq}$ is the corresponding CKM matrix element, the index $i=u,d,s$ denotes the three types of $D$ mesons, $\beta_0 = 11-2n_f /3$ is the first coefficient of the QCD $\beta$-function, and the mass ratio $\rho\equiv m_s^2/m_c^2$. For the perturbative calculation, we have chosen the active quark number $n_f=4$ corresponding to the light quark number $n_l=3$. It has been verified numerically that changing it to $n_f=3$ does not produce a sizable impact. The partoic NLO correction agrees with~\cite{DeFazio:1999ptt,Capdevila:2021vkf}. The NNLO perturbative corrections to $b\to u$ decay width have been calculated in~\cite{vanRitbergen:1999gs,Brucherseifer:2013cu}, a replacement of the active quark flavor number from five to four gives the result for charm decays here.  For the HQET parameters, the hadronic matrix elements of the HQET operators, we accept the convention used by~\cite{Fael:2019umf}, 
\begin{align}
\mu_\pi^2(D_i) &\equiv \langle D_i| \bar c_v(iD)^2 c_v| D_i \rangle /(2m_{D_i}) , \; \mu_G^2(D_i) \equiv \langle D_i| \bar c_v (iD_\alpha)(iD_\beta)(-i\sigma^{\alpha\beta}) c_v| D_i \rangle  /(2m_{D_i}),  \\
\rho_{LS}^3(D_i) &\equiv {1\over2}\langle D_i| \bar c_v \{(iD_\alpha), [iv\cdot D, iD_\beta ]\} (-i\sigma^{\alpha\beta})| D_i \rangle /(2m_{D_i}) , \; 
\rho^3_{D}(D_i) \equiv {1\over2}\langle D_i| \bar c_v  [(iD^\mu), [iv\cdot D, iD_\mu]] c_v | D_i \rangle /(2m_{D_i}). \nonumber 
\end{align}
The dimension-six four-quark operators should theoretically contribute, but they are omitted in practice here due to their vanishing effects under the vacuum insertion approximation~\cite{Bauer:1986bm}. This choice helps avoid an excessive number of free parameters, enabling a successful global fit.

For the initial four electron energy moments, the theoretical results are as follows, 
\begin{align}\label{eq:moments}
\langle E_{e}\rangle_{D_{i}}&=\frac{\hat{\Gamma}_0}{    \Gamma_{D_{i}}}\sum_{q=d,s} \left|V_{cq}\right|^2 m_{c}^6 \left[\frac{3}{10} + \frac{\alpha_{s}}{\pi} a_1^{(1)} + {\alpha_s^2\over \pi^2}a_1^{(2)} -3\rho\delta_{sq} -\frac{1}{2}\frac{\mu_{G}^{2}(D_{i})}{m_{c}^{2}} +\frac{139}{30} \frac{\rho_D^3\left(D_{i}\right)}{m_c^3}+\frac{3}{10} \frac{\rho_{L S}^3\left(D_{i}\right)}{m_c^3}+ ...\right],\\
\langle E_{e}^{2} \rangle_{D_{i}} &=\frac{\hat{\Gamma}_0}{\Gamma_{D_{i}}} \sum_{q=d,s}\left|V_{cq}\right|^2 m_c^7 \left[ \frac{1}{10} + \frac{\alpha_{s}}{\pi} a_2^{(1)}  + {\alpha_s^2\over \pi^2}a_2^{(2)}  - \frac{6}{5} \rho\delta_{sq} + \frac{1}{12}\frac{\mu_{\pi}^2(D_{i})}{ m_c^2} - \frac{11}{60}\frac{\mu_G^2(D_{i})}{ m_c^2} +\frac{17}{6} \frac{\rho_D^3\left(D_{i}\right)}{m_c^3} \right.\nonumber \\
&\qquad\qquad\qquad\qquad\qquad \left. +\frac{7}{30} \frac{\rho_{L S}^3\left(D_{i}\right)}{m_c^3}+ ... \right], \nonumber\\
\left\langle E_e^3\right\rangle_{D_{i}}&=\frac{\hat{\Gamma}_0}{\Gamma_{D_{i}}}\sum_{q=d,s}\left|V_{cq}\right|^2 m_c^8\left[\frac{1}{28} + \frac{\alpha_{s}}{\pi} a_3^{(1)}  + {\alpha_s^2\over \pi^2}a_3^{(2)}  -\frac{1}{2} \rho\delta_{sq} +\frac{1}{14} \frac{\mu_\pi^2\left(D_{i}\right)}{ m_c^2}-\frac{1}{14} \frac{\mu_G^2\left(D_{i}\right)}{ m_c^2}+\frac{223}{140} \frac{\rho_D^3\left(D_{i}\right)}{m_c^3}\right. \nonumber\\
&\qquad\qquad\qquad\qquad\qquad \left.+\frac{1}{7} \frac{\rho_{L S}^3\left(D_{i}\right)}{m_c^3} + ... \right],\nonumber\\
\left\langle E_e^4\right\rangle_{D_i}&=\frac{\hat{\Gamma}_0}{\Gamma_{D_{i}}}\sum_{q=d,s}\left|V_{cq}\right|^2 m_c^9\left[\frac{3}{224} + \frac{\alpha_{s}}{\pi} a_4^{(1)}  + {\alpha_s^2\over \pi^2}a_4^{(2)}  -\frac{3}{14} \rho\delta_{sq} +\frac{3}{64} \frac{\mu_\pi^2\left(D_{i}\right)}{ m_c^2}-\frac{13}{448} \frac{\mu_G^2\left(D_{i}\right)}{ m_c^2} +\frac{481}{560} \frac{\rho_D^3\left(D_i\right)}{m_c^3}\right. \nonumber\\
&\qquad\qquad\qquad\qquad\qquad \left.+\frac{9}{112} \frac{\rho_{L S}^3\left(D_i\right)}{m_c^3}+ ... \right],\nonumber
\end{align}
where the NLO and NNLO coefficients are given by 
\begin{align}
a^{(1)}_1 &= {1093 - 180\pi^2\over 900 }, \;
a^{(1)}_2 = \frac{4243 - 720 \pi^2}{{10800}}, \; 
a^{(1)}_3 = \frac{144037-25200 \pi^2}{{1058400}}, \; 
a^{(1)}_4 = \frac{69827-12600 \pi^2}{{1411200}},   \\
a^{(2)}_1 &= {\beta_0\over 4}a^{(1)}_1\ell -7.70077, \; 
a^{(2)}_2 = {\beta_0\over 4}a^{(1)}_2\ell -2.77835, \; 
a^{(2)}_3 = {\beta_0\over 4}a^{(1)}_3\ell -1.06371 , \; 
a^{(2)}_4 = {\beta_0\over 4}a^{(1)}_4\ell -0.42438 , \nonumber
\end{align}
with $\ell\equiv \log(\mu^2/m_c^2)$. The NLO results $a^{(1)}_{1-4}$ are obtained by phase space integration of the analytical differential decay widths given by~\cite{Aquila:2005hq}. The numerical NNLO results $a^{(2)}_{1-4}$ are provided by authors of~\cite{Chen:2023osm}.
Our results for the NLO corrections to the first two electron energy moments are consistent with~\cite{Kamenik:2011zz}, and first four ones at LO align with~\cite{Fael:2019umf}. 

The aforementioned results are highly sensitive to the charm quark mass $m_{c}$, given that they are proportional to high powers of $m_{c}$. A proper choice of the charm mass, such as its scheme, is essential for precise theoretical predictions and hence the extraction of non-perturbative parameters. While the perturbative calculations mentioned above rely on the pole mass, this choice is deemed inappropriate due to the renormalon ambiguity. To circumvent this issue, an appropriate short-distance mass is required. The kinetic mass scheme has been effectively utilized in semileptonic $B$ decays, utilizing a cutoff scale of 1 GeV~\cite{Alberti:2014yda,Gambino:2016jkc}. However, this approach does not result in a convergent expansion in the case of charm~\cite{Boushmelev:2023kmf}. In the following, we will thus consider two other charm mass schemes:\footnote{An alternative method for treating the quark mass within the heavy quark expansion is proposed by~\cite{Boushmelev:2023kmf}.}
\begin{itemize}
    \item The $\ov{\mathrm{MS}}$ mass scheme: the pole mass $m_c$ is expressed in terms of the $\ov{\mathrm{MS}}$ mass $\ov{m}_c(\mu)$ as~\cite{Gray:1990yh,Broadhurst:1991fy,Fleischer:1998dw,Melnikov:2000qh} 
\begin{align}\label{eq:msbarmass}
m_c=\ov{m}_c\left(\mu \right)\Big[ &1 + \frac{\alpha_s\left(\mu\right)}{\pi} \left(\frac{4}{3} + \log\left({\mu^2\over \ov{m}_c^2}\right) \right) 
+ \frac{\alpha_s^2\left(\mu\right)}{\pi^2}\frac{1}{288} \Big( 112 \pi ^2+2905+16
   \pi ^2 \log (4)  -48 \zeta (3)   \\ 
   & -12 (2n_f-45) \log
   ^2\left(\frac{\mu^2}{\ov{m}_c^2}\right) -4 (26
   n_f-519) \log
   \left(\frac{\mu^2}{\ov{m}_c^2}\right)-2 \left(71+8
   \pi ^2\right) n_f\Big)+ \mathcal{O}(\alpha_s^3) \Big]\; . \nonumber
\end{align}
\item The 1S mass scheme: the pole mass $m_c$ is expressed in terms of the 1S mass $m_{c,\mathrm{1S}}$ as~\cite{Hoang:1998ng,Hoang:1998hm, Hoang:1999zc} 
\begin{equation}\label{eq:1smass}
m_c =m_{c,\mathrm{1S}}+ m_{c,\mathrm{1S}} \frac{ \alpha_s(\mu)^2 C_F^2 }{8}\left\{ 1 + {\alpha_s\over\pi} \left[ \left( - \log\left(\alpha_s(\mu) m_{c,\mathrm{1S}} C_F/\mu\right)+  {11\over6}\right)\beta_0 -4 + {\pi\over 8}C_F \alpha_s \right] + ... \right\}  .
\end{equation}
\end{itemize}

For the $\ov{\mathrm{MS}}$ mass scheme, we substitute the pole mass in \eqref{eq:decaywidth} and \eqref{eq:moments} with \eqref{eq:msbarmass} and expand consistently up to order $\alpha_{s}^2$. The formulas are listed in~\eqref{eq:msbar}.

For the 1S mass scheme, we substitute the pole mass in \eqref{eq:decaywidth} and \eqref{eq:moments} with \eqref{eq:1smass} and also perform the expansion up to order $\alpha_{s}^2$. Note that the correction within the 1S scheme \eqref{eq:1smass} in fact starts at order $\alpha_{s}^2$ which however is still considered to be an NLO (not NNLO) eﬀect, and the order $\alpha_{s}^3$ is considered to be an NNLO (not NN\textcolor{blue}{N}LO) eﬀect~\cite{Hoang:1998ng}. The formulas are listed in~\eqref{eq:1s}.

A key issue for the charm decay is the convergence of the perturbative expansion, as the strong coupling at the charm scale $\alpha_s\left(\left(\bar{m}_c\right)\approx 1.27 \mathrm{GeV} \right) \approx 0.387$ is not quite small. Indeed, the results in the pole mass scheme show very poor convergence. We have tested that for the decay width, the $\alpha_s$ expansion $1 -0.768104 \alpha_s -2.37521 \alpha_s^2 -10.7295 \alpha_s^3$ even becoming negative when we include the next-to-next-to-next-to-leading order (N$^3$LO) result provided by authors of~\cite{Chen:2023osm}. However, if we transform to the 1S mass scheme, it turns out that the NLO, NNLO and N$^3$LO  corrections modify the LO decay width by -13.1\%, -4.8\% and +1.9\%, respectively. The results including the N$^3$LO  corrections indicates the validity of the perturbative expansion in inclusive charm decays, and the 1S mass scheme would be a favorable choice.

\section{EXPERIMENTAL DATA}

The inclusive $D^+,D^0,D_s^+$ decay widths and electron energy spectra have been measured by the CLEO collaboration~\cite{CLEO:2009uah}, and the $D_s^+$ results were updated by BESIII~\cite{BESIII:2021duu} with improved precision. Therefore, in this analysis we accept the CLEO data for $D^+,D^0$ and the BESIII data for $D_s^+$.

Both CLEO and BESIII set a lower cut on the electron momentum (equal to the electron energy $E_e$ with neglecting the electron mass) in the laboratory frame of $p_{e}>0.2\GeV$~\cite{CLEO:2009uah,BESIII:2021duu} in the measurements of the electron energy spectra. To match our theoretical formulas for the energy moments, which requires integration over the entire phase space, the spectra towards $p_{e}=0$ need extrapolation from known data. Following the procedure used in~\cite{Gambino:2010jz} supposing a well feature of OPE in the $p_e \approx 0$ region, we fit the first four measured bins to $d \Gamma / d x=a x^2(1+b x)(1-x)$, with $x=2 p_e / m_c$. Then, we perform a Monto Carlo simulation of the electron energy distribution based on the binned measurements. In the simulation, we assume that the event number in each bin obeys the normal distribution and that the statistical uncertainties between different bins are not correlated while the systematic uncertainties are fully correlated. In addition, the electron energy in each bin is assumed to be flatly distributed. With 200,000 simulated data samples each for $D^+,D^0,D_s^+$, we compute the following branching ratios,
\begin{equation}
    B(D_{s}\to X e^+\nu_{e})=0.0631(14)\quad\quad\quad\quad B(D^{0}\to X e^+\nu_{e})=0.0636(15)\quad\quad\quad\quad B(D^{+}\to X e^+\nu_{e})=0.1602(32)
    \label{Br}
\end{equation}
which are fully compatible with the reported results by CLEO and BESIII~\cite{CLEO:2009uah,BESIII:2021duu}. The first four electron energy moments are also obtained, as 
\begin{equation}
\begin{aligned}
\left\langle E_e\right\rangle_{\rm lab}^{D_{s}}=0.451(6) \mathrm{GeV}, & \quad\left\langle E_e^2\right\rangle_{\rm lab}^{D_{s}}=0.239(5) \mathrm{GeV}^2&\left\langle E_e^3\right\rangle_{\rm lab}^{D_{s}}=0.142(4) \mathrm{GeV}, &\quad \left\langle E_e^4\right\rangle_{\rm lab}^{D_{s}}=0.092(3) \mathrm{GeV}^2 \; , \\
\left\langle E_e\right\rangle_{\rm lab}^{D^0}=0.467(5) \mathrm{GeV}, &\quad\left\langle E_e^2\right\rangle_{\rm lab}^{D^0}=0.250(5) \mathrm{GeV}^2 &\left\langle E_e^3\right\rangle_{\rm lab}^{D^0}=0.146(4) \mathrm{GeV}, & \quad\left\langle E_e^4\right\rangle_{\rm lab}^{D^0}=0.092(3) \mathrm{GeV}^2 \; , \\
\left\langle E_e\right\rangle_{\rm lab}^{D^{+}}=0.459(4) \mathrm{GeV}, & \quad\left\langle E_e^2\right\rangle_{\rm lab}^{D^{+}}=0.242(4) \mathrm{GeV}^2&\left\langle E_e^3\right\rangle_{\rm lab}^{D^{+}}=0.140(3) \mathrm{GeV}, &\quad \left\langle E_e^4\right\rangle_{\rm lab}^{D^{+}}=0.087(3) \mathrm{GeV}^2 \;, \\
\end{aligned}
\label{momentLab}
\end{equation}
with the first two moments of $D^{+,0}$ compatible with~\cite{Gambino:2010jz}. Note that these moment values are obtained in the laboratory frame, which still need to be boosted to the rest frames of the $D$ mesons. 

The energy of the final state electron in the laboratory frame is given by $E_{\mathrm{e}}^{\prime}=\gamma E_{\mathrm{e}}(1-\beta \cos \theta)$, where $E_{\mathrm{e}}$ is the electron energy in the $D$ meson rest frame, $\beta,\gamma=1/\sqrt{1-\beta^2}$ are the Lorentz boost factors of the initial $D$ mesons, and $\theta$ is the angle between the electron momentum in the $D$ rest frame and the $D$ meson momentum in the laboratory frame. Because the $D$ meson is spin zero, $\theta$ is flatly distributed. We thus obtain the relations between the electron energy moments in the laboratory frames and in $D$ rest frames, as 
\begin{equation}\label{eq:boost}
  \left\langle E_e\right\rangle_{\rm lab}=\gamma\left\langle E_e\right\rangle , \;  \left\langle E_e^2\right\rangle_{\rm lab}= (1 + {\beta^2\over3}) \gamma^2\left\langle E_e^2\right\rangle, \; \left\langle E_e^3\right\rangle_{\rm lab}=(1 + \beta^2) \gamma^3\left\langle E_e^3\right\rangle,\; \left\langle E_e^4\right\rangle_{\rm lab}= (1 + 2\beta^2 + {\beta^4\over5}) \gamma^4\left\langle E_e^4\right\rangle . 
\end{equation}
For the $D^{+,0}$ mesons at CLEO, $\gamma=1.009,1.012$~\cite{Gambino:2010jz}. In the case of the $D_s^+$ meson at BESIII, data of multiple collision energy points are collected~\cite{BESIII:2021duu}, so we use the weight averaged value $\gamma=1.027$.
Applying these boost factors to \eqref{eq:boost}, we obtain the four moments in the $D$ meson rest frame,
\begin{equation}
\begin{aligned}
\left\langle E_e\right\rangle_{\rm exp}^{D_{s}}=0.439(5) \mathrm{GeV}, & \quad\left\langle E_e^2\right\rangle_{\rm exp}^{D_{s}}=0.223(5) \mathrm{GeV}^2, &\left\langle E_e^3\right\rangle_{\rm exp}^{D_{s}}=0.124(4) \mathrm{GeV}^3, &\quad \left\langle E_e^4\right\rangle_{\rm exp}^{D_{s}}=0.074(3) \mathrm{GeV}^4, \\
\left\langle E_e\right\rangle_{\rm exp}^{D^0}=0.462(5) \mathrm{GeV}, &\quad\left\langle E_e^2\right\rangle_{\rm exp}^{D^0}=0.242(5) \mathrm{GeV}^2,  &\left\langle E_e^3\right\rangle_{\rm exp}^{D^0}=0.138(4) \mathrm{GeV}^3, & \quad\left\langle E_e^4\right\rangle_{\rm exp}^{D^0}=0.084(3) \mathrm{GeV}^4, \\
\left\langle E_e\right\rangle_{\rm exp}^{D^{+}}=0.455(4) \mathrm{GeV}, & \quad\left\langle E_e^2\right\rangle_{\rm exp}^{D^{+}}=0.236(4) \mathrm{GeV}^2, &\left\langle E_e^3\right\rangle_{\rm exp}^{D^{+}}=0.134(3) \mathrm{GeV}^3, &\quad \left\langle E_e^4\right\rangle_{\rm exp}^{D^{+}}=0.081(3) \mathrm{GeV}^4. \\
\end{aligned}
\label{momentRest}
\end{equation}
The correlation matrices of the branching ratios and the electron energy moments are detailed in~\eqref{eq:cormoments}. It is evident that the electron energy moments exhibit high correlations, leading to nearly singular correlation matrices. Consequently, in our practical analysis, we employ linear combinations of the moments, specifically, 
\begin{align}
\left\langle E_e\right\rangle, \;\left\langle E_e^2\right\rangle_{\rm center}\equiv \left\langle (E_e - \langle E_e\rangle)^2 \right\rangle , \; \left\langle E_e^3\right\rangle_{\rm center}\equiv\left\langle (E_e - \langle E_e\rangle)^3 \right\rangle , \; \left\langle E_e^4\right\rangle_{\rm center}\equiv\left\langle (E_e - \langle E_e\rangle)^4 \right\rangle .
\end{align}
Their central values and uncertainties are 
\begin{equation}
\begin{aligned}
\left\langle E_e^2 \right\rangle_{\text{exp,center}}^{D_{s}} &= 0.0297(13) \, \mathrm{GeV}^2, & \left\langle E_e^3 \right\rangle_{\text{exp,center}}^{D_{s}} &= 0.0004(4) \, \mathrm{GeV}^3, & \quad \left\langle E_e^4 \right\rangle_{\text{exp,center}}^{D_{s}} &= 0.0021(2) \, \mathrm{GeV}^4, \\
\left\langle E_e^2 \right\rangle_{\text{exp,center}}^{D^0} &= 0.0287(12) \, \mathrm{GeV}^2, & \left\langle E_e^3 \right\rangle_{\text{exp,center}}^{D^0} &= -0.0001(3) \, \mathrm{GeV}^3, & \quad \left\langle E_e^4 \right\rangle_{\text{exp,center}}^{D^0} &= 0.0019(1) \, \mathrm{GeV}^4, \\
\left\langle E_e^2 \right\rangle_{\text{exp,center}}^{D^{+}} &= 0.0291(11) \, \mathrm{GeV}^2, & \left\langle E_e^3 \right\rangle_{\text{exp,center}}^{D^{+}} &= -0.0002(3) \, \mathrm{GeV}^3, & \quad \left\langle E_e^4 \right\rangle_{\text{exp,center}}^{D^{+}} &= 0.0019(1) \, \mathrm{GeV}^4 , 
\end{aligned}
\label{centermomentRest}
\end{equation}
with the correlation matrices also given in~\eqref{eq:corcenmoments}.

\section{Phenomenological Analysis}

In order to determine the HQET parameters in inclusive charm decays from experimental data, we employ a fitting procedure by matching the theoretical formulas for the decay widths~\eqref{eq:decaywidth} and the electron energy moments~\eqref{eq:moments} of $\left\{D^0, D^{+}, D_s\right\} \to X_{s,d}\ell \bar{\nu}$ with the corresponding experimental measurements.
We perform our analysis up to order $\Lambda_{\mathrm{QCD}}^3 / m_c^3$ in the heavy quark expansion to determine the following HQET parameters,
\begin{equation}
    \mu_{\pi}^2(D^{0,+}),\quad \mu_{G}^{2}(D^{0,+}),\quad \rho_{D}^3(D^{0,+}),\quad \rho_{
    LS
    }^3(D^{0,+}) ,\quad\mu_{\pi}^2(D_{s}),\quad \mu_{G}^{2}(D_{s}),\quad \rho_{D}^3(D_{s}),\quad \rho_{
    LS
    }^3(D_{s}) ,
\end{equation}
where the parameters for $D^+$ and $D^0$ are supposed to be identical owing to the isospin symmetry. Notably, these parameters will play a crucial role in the computation of various observables, including $D$ meson lifetimes, rare inclusive $D$ meson decay rates, and even inclusive $B$ meson decay rates. As for the four-quark operators which start contributing from the order $\Lambda_{\mathrm{QCD}}^3 / m_c^3$, we find that their contributions are negligible under the  vacuum insertion approximation~\cite{Bauer:1986bm} at this level. 

To assess the convergence of the heavy quark expansion, we conduct the fitting process under two distinct scenarios. In Scenario 1, we only consider the contributions from dimension-five operators to the observables. In Scenario 2, we also include the contributions from dimention-six operators. The experimental data used in both the schemes include the inclusive decay widths~\eqref{Br}, the first electron energy moments~\eqref{momentRest}, the second, third and fourth order  electron energy center moments~\eqref{centermomentRest}. We take into account the correlation among the various observables, and the correlation matrix is provided in~\eqref{eq:corcenmoments}. As previously mentioned, the theoretical outcomes are highly dependent on the choice of the charm mass. For each fitting scenario, we explore both the $\ov{\mathrm{MS}}$ mass and 1S mass schemes.

In the fit, we use the $2+1+1$ Lattice QCD FLAG averages for the strange quark mass, $\overline{m}_s(2 \mathrm{GeV})=(93.44 \pm 0.68)$ MeV~\cite{FlavourLatticeAveragingGroup:2019iem}. For the charm mass, we use $\ov{m}_c(\ov{m}_c) = 1.27\pm0.02$ GeV and $m_{c,\mathrm{1S}} = 1.55$ GeV~\cite{ParticleDataGroup:2024cfk}. For the strong coupling constant, we use $\alpha_{s}(\ov{m}_{c})=0.387$ obtained in~\cite{King:2021xqp} using the RunDec package~\cite{Herren:2017osy}. For $\alpha_s(\mu)$ and $\ov{m}_c(\mu)$, we consider the running energy scale $\mu$ from 1 GeV to $2\ov{m}_{c}(\ov{m}_{c})$. We use the following input values for the remaining parameters in the theoretical formulas~\cite{ParticleDataGroup:2024cfk}, $G_F=1.1663788\times10^{-5}$, $|V_{cs}|=0.975 \pm 0.006$, and $|V_{cd}|=0.221 \pm 0.004$.

\begin{table}[h!]
\centering
\begin{tabular}{c|c|l|llll} 
\hline
  $\mathrm{\overline{MS}}$ scheme & $\chi^2/{\rm d.o.f.}$ & $D_{i}$&$\mu_{\pi}^{2}$/$\mathrm{GeV^2}$&$\mu_{G}^{2}$/$\mathrm{GeV^2}$&$\mathrm{\rho_{D}^{3}}$/$\mathrm{GeV^3}$ &$\rho_{LS}^{3}$/$\mathrm{GeV^3}$ \\ 
  \hline
 \multirow{2}{*}{Scenario 1} & \multirow{2}{*}{ {4.5}}  & $D^{0,+}$& {$0.09\pm0.01$}& {$0.27 \pm0.14$}&-&-\\
 &&$D_{s}$& {$0.09\pm 0.02$}& {$0.39\pm 0.12$}&-&-\\
  \hline
  \multirow{2}{*}{Scenario 2}  &  \multirow{2}{*}{ {2.1}}  & $D^{0,+}$ & {$0.11\pm0.02$}& {$0.26\pm0.14$}& {$-0.002\pm 0.002$}& {$0.003\pm 0.002$}\\
  &&$D_{s}$& {$0.12\pm0.02$}& {$0.38 \pm0.13$}& {$-0.003\pm0.002$}& {$0.005\pm 0.002$}\\  \hline
\end{tabular}
\caption{The $\chi^2$ fitting results in the $\mathrm{\overline{MS}}$ mass scheme. Scenario 1 excludes the dimension-six operator contributions while Scenario 2 includes them. The $\chi^2$/d.o.f. in the fit, along with the central values and uncertainties for the HQET parameters, are displayed.}
\label{tab:fitting result on msbar mass}
\end{table}

\begin{table}[h]
\centering
\begin{tabular}{c|c|l|llll} 
\hline
 1S scheme & $\chi^2/{\rm d.o.f.}$ & $D_{i}$ & $\mu_{\pi}^{2}$/$\mathrm{GeV^2}$&$\mu_{G}^{2}$/$\mathrm{GeV^2}$&$\mathrm{\rho_{D}^{3}}$/$\mathrm{GeV^3}$ &$\rho_{LS}^{3}$/$\mathrm{GeV^3}$ \\ 
  \hline
 \multirow{2}{*}{Scenario 1} & \multirow{2}{*}{ {4.9}}  & $D^{0,+}$& {$0.04\pm0.01$}& {$0.33\pm 0.02$}&-&-\\
 &&$D_{s}$& {$0.06\pm 0.02$}& {$0.44\pm 0.02$}&-&-\\
  \hline
  \multirow{2}{*}{Scenario 2}  &  \multirow{2}{*}{ {$0.33$}}  & $D^{0,+}$& {$0.09\pm0.02$}& {$0.32\pm0.02$}& {$-0.003\pm0.002$}& {$0.004\pm 0.002$} \\
  &&$D_{s}$& {$0.11\pm0.02$}& {$0.43\pm 0.02$}& {$-0.004\pm 0.002$}& {$0.005\pm0.002$}\\  \hline
\end{tabular}
\caption{Same as Table~\ref{tab:fitting result on msbar mass} except for the 1S scheme.}
\label{tab:fitting result on 1S mass}
\end{table}

The fitting results for in the $\ov{\rm MS}$ mass scheme and in the 1S mass scheme are summarized in Table~\ref{tab:fitting result on msbar mass} and  Table~\ref{tab:fitting result on 1S mass}, respectively. The $\chi^2$'s per degree of freedom ($\chi^2$/d.o.f.) along with the central values of the HQET parameters and their uncertainties are provided. It is evident that the inclusion of dimension-six operator contributions significantly enhances the fitting quality in both mass schemes, with notably smaller $\chi^2$/d.o.f. values observed in Scenario 2. Particularly, Scenario 2 in the 1S scheme exhibits a $\chi^2$/d.o.f. less than 1, indicating a successful global fit. The extracted values for $\mu^2_\pi$ and $\mu_G^2$ remain relatively stable across both scenarios, suggesting a reliable heavy quark expansion in semi-leptonic inclusive charm decays. Given the favorable convergence of the 1S scheme in the $\alpha_s$ expansion and its superior fitting performance, we adopt the results from ``Scenario 2, 1S scheme" as our primary outcomes, with their discrepancies from ``Scenario 1, 1S scheme" considered as systematic uncertainties arising from unknown power corrections. They can be quoted as 
\begin{align}
\mu^2_\pi(D^{0,+}) &= (0.09\pm 0.05) \mathrm{GeV}^2, \qquad \qquad
\mu^2_\pi(D^{+}_s) = (0.11\pm 0.05) \mathrm{GeV}^2,  \\
\mu^2_G(D^{0,+}) &= (0.32\pm 0.02) \mathrm{GeV}^2, \qquad \qquad 
\mu^2_G(D^{+}_s) = (0.43\pm 0.02) \mathrm{GeV}^2, \nonumber \\
\rho_D^3(D^{0,+}) &= (-0.003\pm 0.002) \mathrm{GeV}^3, \qquad\ 
\rho_D^3(D^{+}_s) = (-0.004\pm 0.002) \mathrm{GeV}^3, \nonumber \\
\rho_{LS}^3(D^{0,+}) &= (0.004\pm 0.002) \mathrm{GeV}^3, \qquad \ \ \ 
\rho_{LS}^3(D^{+}_s) = (0.005\pm 0.002) \mathrm{GeV}^3 , \nonumber 
\end{align}
which are exactly the values that are presented in the abstract. 

To further validate the fit in ``Scenario 2, 1S scheme", we compute the decay widths and electron energy moments in the inclusive decays utilizing the extracted HQET parameters. The results are presented in Table~\ref{tab:obs on 1S}. It is evident that they are in excellent agreement with the experimental data. 

\begin{table}[h]
\centering
\begin{tabular}{c|l|l|l|l|l}
\hline
&$\Gamma_{sl}/\GeV$&$\langle E_{e}\rangle/\GeV$ &$\langle E_{e}^2\rangle/\GeV^2$&$\langle E_{e}^3\rangle/\GeV^3$&$\langle E_{e}^4\rangle/\GeV^4$\\
 \hline
 $D^{0,+}$&$(1.023 \pm 0.016) \times 10^{-13}$&$0.457\pm0.002$&$0.238\pm 0.003$&$0.135\pm 0.002$&$0.082\pm 0.002$\\
$D_s$ &  $(0.824\pm 0.019) \times 10^{-13}$ & $0.443\pm 0.003$ & $0.226\pm 0.004$ & $0.126\pm 0.003$ & $0.076\pm 0.003$ \\
\hline
\end{tabular}
\caption{The predictions for the inclusive $D$ decay widths and the electron energy moments in ``1S scheme, Scenario 2".}
\label{tab:obs on 1S}
\end{table}

\section{Summary and prospect}
For the first time, we have determined the HQET parameters $\mu_\pi^2$, $\mu_G^2$, $\rho_D^3$, and $\rho_{LS}^3$ in inclusive $D$ meson decays from data in a model-independent manner. These parameters will serve as essential inputs to the calculation of various observables like $D$ meson lifetimes, rare inclusive $D$ meson decay rates, inclusive $B$ meson decay rates, and so on. They will also be subject to scrutiny from, for example, lattice calculations~\cite{Gambino:2017vkx,Gambino:2020crt,Gambino:2022dvu,Nefediev:2024mjk} to test the validity of heavy quark expansion in inclusive charm decays. It is important to note that the current state of affairs is still unsatisfactory, given the significant uncertainties in the extracted $\mu_\pi^2$ and $\mu_G^2$. We anticipate that collaborative efforts from both the theoretical and experimental communities will lead to a reduction in these uncertainties.

From a theoretical perspective, higher-order radiative corrections are necessary to match the experimental precision. Of utmost importance are the N$^3$LO corrections to the leading power contributions to the electron energy moments~\cite{Chen:2023osm,Chen:2023dsi,Yan:2024hbz,Fael:2020tow,Fael:2023tcv,Fael:2024vko}. In terms of power corrections, in Scenario 1, only dimension-five operator contributions were included; even in Scenario 2, dimension-six operators were included in the vacuum insertion approximation of four-quark operators, resulting in their contribution being negligible. A more comprehensive set of operators, including up to dimension-seven, is anticipated in future work to consistently extract the matrix elements of these operators. This, however, necessitates the measurement of additional observables in experiments, which will be further discussed in the following paragraph. In addition, the determination of the short-distance coefficients of the power corrections to the electron energy moments also awaits results with a higher power expansion in $\alpha_s$~\cite{Capdevila:2021vkf}.

From an experimental perspective, the BESIII and upcoming tau-charm factories~\cite{Achasov:2023gey} are expected to offer more precise measurements of the total and differential decay rates, which are essential for better determination of the HQET parameters. Furthermore, we suggest that experiments incorporate the following enhancements compared to prior investigations:
\begin{itemize}
    \item Present the results for the differential decay rates in the rest frame of $D$ mesons instead of in the laboratory frame. It allows theorists to directly utilize the results without the need for Lorentz boost relations, which are only valid for the electron energy moments in the entire phase space. Consequently, the electron energy moments with a lower energy cut can be obtained, providing more experimental observables for HQET parameter determination.
    \item Present the measured electron energy moments directly. Obtaining the electron energy moments from a binned electron energy spectrum requires assuming a specific distribution within each bin, introducing unnecessary uncertainties. Experimental results derived directly from events can mitigate such uncertainties and are thus more accurate.
    \item Utilize the information from the inclusive hadronic system $X$ and reconstruct the invariant mass squares of the lepton pair ($q^2$) and the hadronic system ($M_X$). The results for $q^2$ moments, in particular, will enhance the determination of matrix elements of HQET operators with reparametrization invariance. By including sufficient observables such as the $E_e$, $q^2$ and $M_X$ moments with various cuts, it becomes feasible to determine the matrix elements of dimension-seven operators.
    \item Distinguish between $D\to eX$ into $D\to eX_s$ and $D\to eX_d$. This differentiation will enable the inclusive determination of $V_{cs}$ and $V_{cd}$, allowing for a comparison with exclusive values to test the CKM mechanism.
    \item Perform measurements for muonic channels $D\to \mu X$. These measurements can complement the electronic channels and, importantly, the $\mu/e$ ratios for all observables can significantly reduce theoretical uncertainties, rendering them precise test observables for the standard model.
\end{itemize}

\section*{Acknowledgements}
We wish to thank Matteo Fael and Wen-Jie Song for enlightening discussions on calculations of electronic energy spectrum, to Ying-Ao Tang for suggestions on the extraction of electronic energy moments, to Bo-Nan Zhang, Ji-Xin Yu and Yong Zheng for some hints on numerical testing, to Guo-He Yang for collaborating at early stages of this work, and especially to Dong Xiao and Fu-Sheng Yu for their inspiring discussions on various experimental and theoretical aspects, respectively. Special thanks go to Long Chen and Yan-Qing Ma for providing the numerical NNLO corrections to the partonic decay widths and the corresponding electron energy moments. This work is supported by Natural Science Foundation of China under grant No.~12375086 and No.~12335003.

\begin{appendix}
\section{Formulas for the $\overline{\mathrm{MS}}$ mass scheme and 1S mass scheme}\label{app:massscheme}

The theoretical formulas for the leading power contributions to the decay widths and the electron energy moments in the $\ov{\mathrm{MS}}$ scheme are given by 
\begin{align}\label{eq:msbar}
\Gamma_{D_{i}}&=\hat{\Gamma}_0 \sum_{q=d,s}\left|V_{cq}\right|^2 \ov{m}_c^5(\mu) \left[ 1 +\frac{\alpha_s}{\pi} \left(5 \bar{\ell}+4.2536 \right) + \frac{\alpha_s^2}{\pi^2}\left(\frac{425}{24} \bar{\ell}^2+38.3935 \bar{\ell}+29.8447 \right)  \right]  ,\\ 
\langle E_{e}\rangle_{D_{i}}&=\frac{\hat{\Gamma}_0}{    \Gamma_{D_{i}}}\sum_{q=d,s} \left|V_{cq}\right|^2 \ov{m}_c^6(\mu) \left[\frac{3}{10} + \frac{\alpha_{s}}{\pi} \left(\frac{9}{5} \bar{\ell}+1.64052\right) + {\alpha_s^2\over \pi^2}\left(\frac{291}{40} \bar{\ell}^2+16.2359 \bar{\ell}+12.7981\right) \right],\nonumber \\
\langle E_{e}^{2} \rangle_{D_{i}} &=\frac{\hat{\Gamma}_0}{\Gamma_{D_{i}}} \sum_{q=d,s}\left|V_{cq}\right|^2 \ov{m}_c^7(\mu) \left[ \frac{1}{10} + \frac{\alpha_{s}}{\pi} \left(\frac{7}{10} \bar{\ell}+0.66823\right)  + {\alpha_s^2\over \pi^2}\left(\frac{763}{240} \bar{\ell}^2+7.2267 \bar{\ell}+5.70419\right)  \right], \nonumber \\
\left\langle E_e^3\right\rangle_{D_{i}}&=\frac{\hat{\Gamma}_0}{\Gamma_{D_{i}}}\sum_{q=d,s}\left|V_{cq}\right|^2 \ov{m}_c^8(\mu) \left[\frac{1}{28} + \frac{\alpha_{s}}{\pi} \left(\frac{2}{7} \bar{\ell}+0.282051\right)  + {\alpha_s^2\over \pi^2}\left(\frac{121}{84} \bar{\ell}^2+3.31624 \bar{\ell}+2.60749\right)   \right],\nonumber \\
\left\langle E_e^4\right\rangle_{D_i}&=\frac{\hat{\Gamma}_0}{\Gamma_{D_{i}}}\sum_{q=d,s}\left|V_{cq}\right|^2 \ov{m}_c^9(\mu) \left[\frac{3}{224} + \frac{\alpha_{s}}{\pi} \left(\frac{27}{224} \bar{\ell}+0.122073\right)  + {\alpha_s^2\over \pi^2}\left(\frac{171}{256} \bar{\ell}^2+1.5522 \bar{\ell}+1.21291\right)  \right],\nonumber 
\end{align}
where $\bar{\ell}\equiv \log\left({\mu^2}/{\ov{m}_c^2}\right)$. 

The theoretical formulas for the leading power contributions to the decay widths and the electron energy moments in the 1S scheme are given by
\begin{align}\label{eq:1s}
\Gamma_{D_{i}}&=\hat{\Gamma}_0 \sum_{q=d,s}\left|V_{cq}\right|^2 m_{c,\mathrm{1S}}^5 \Big[1+\epsilon  \left(\frac{10 \alpha _s^2}{9}-\frac{2 \pi \alpha _s}{3}+\frac{25 \alpha _s}{6 \pi }\right)\\
   & \qquad\qquad \qquad\qquad \qquad +\epsilon
   ^2 \left(2.9473 \ell_{\alpha} \alpha _s^3-0.50936
   \ell_{\text{1s}} \alpha _s^2+0.74074 \alpha _s^4+3.1352 \alpha
   _s^3-2.3752 \alpha _s^2\right) \Big] \; , \nonumber  \\
\langle E_{e}\rangle_{D_{i}}&=\frac{\hat{\Gamma}_0}{    \Gamma_{D_{i}}}\sum_{q=d,s} \left|V_{cq}\right|^2 m_{c,\mathrm{1S}}^6 \Big[ \frac{3}{10}+\epsilon  \left(\frac{2 \alpha_s^2}{5}-\frac{\pi  \alpha _s}{5}+\frac{1093 \alpha_s}{900 \pi }\right)\nonumber \\
   & \qquad\qquad \qquad\qquad \qquad +\epsilon ^2 \left(-0.16031 \ell_{\text{1s}} \alpha_s^2+1.061 \ell_{\alpha} \alpha_s^3+0.31111 \alpha _s^4+1.1136 \alpha _s^3-0.78025 \alpha_s^2\right)  \Big],\nonumber \\
\langle E_{e}^{2} \rangle_{D_{i}} &=\frac{\hat{\Gamma}_0}{\Gamma_{D_{i}}} \sum_{q=d,s}\left|V_{cq}\right|^2 m_{c,\mathrm{1S}}^7 \Big[ \frac{1}{10}+\epsilon  \left(\frac{7 \alpha_s^2}{45}-\frac{\pi  \alpha _s}{15}+\frac{4243 \alpha_s}{10800 \pi }\right)\nonumber \\
   & \qquad\qquad \qquad\qquad \qquad +\epsilon ^2 \left(-0.055960 \ell_{\text{1s}} \alpha_s^2+0.41262 \ell_{\alpha} \alpha_s^3+0.13827 \alpha_s^4+0.42715 \alpha _s^3-0.28151 \alpha _s^2\right)  \Big], \nonumber \\
\left\langle E_e^3\right\rangle_{D_{i}}&=\frac{\hat{\Gamma}_0}{\Gamma_{D_{i}}}\sum_{q=d,s}\left|V_{cq}\right|^2 m_{c,\mathrm{1S}}^8 \Big[\frac{1}{28}+\epsilon  \left(\frac{4 \alpha_s^2}{63}-\frac{\pi  \alpha _s}{42}+\frac{144037 \alpha_s}{1058400 \pi }\right)\nonumber \\
   & \qquad\qquad \qquad\qquad \qquad +\epsilon ^2 \left(-0.020877 \ell_{\text{1s}} \alpha_s^2+0.16842 \ell_{\alpha} \alpha_s^3+0.063492 \alpha_s^4+0.17196 \alpha_s^3-0.10778 \alpha_s^2\right)   \Big],\nonumber \\
\left\langle E_e^4\right\rangle_{D_i}&=\frac{\hat{\Gamma}_0}{\Gamma_{D_{i}}}\sum_{q=d,s}\left|V_{cq}\right|^2 m_{c,\mathrm{1S}}^9 \Big[ \frac{3}{224}+\epsilon  \left(\frac{3 \alpha_s^2}{112}-\frac{\pi  \alpha _s}{112}+\frac{69827 \alpha_s}{1411200 \pi }\right)\nonumber \\
   & \qquad\qquad \qquad\qquad \qquad +\epsilon^2 \left(-0.0081565 \ell_{\text{1s}} \alpha_s^2+0.071051 \ell_{\alpha} \alpha_s^3+0.029762 \alpha_s^4+0.071557 \alpha_s^3-0.042999 \alpha_s^2\right)  \Big],\nonumber 
\end{align}
where $\ell_\alpha\equiv -\log({\alpha_s(\mu) m_{c,\mathrm{1S}} C_F}/\mu)$, $\ell_{\text{1S}}\equiv \log(\mu^2/m_{c,\mathrm{1S}}^2)$ and the variable $\epsilon=1$ is used for the modified expansion~\cite{Hoang:1998hm}.

\section{Correlation matrices}
There are two sets of observables. The first one includes the decay width and the first four electron energy moments. Their correlation matrices for the $D^0$, $D^+$ and $D_s^+$ decays are given by  
\begin{align}\label{eq:cormoments}
{\rm Cor}(D^0) &= \left(\begin{array}{ccccc}
1 . & -0.0764818 & -0.0682164 & -0.0584038 & -0.049709 \\
-0.0764818 & 1 . & 0.964628 & 0.888372 & 0.799743 \\
-0.0682164 & 0.964628 & 1 . & 0.976053 & 0.921246 \\
-0.0584038 & 0.888372 & 0.976053 & 1 . & 0.982859 \\
-0.049709 & 0.799743 & 0.921246 & 0.982859 & 1 .
\end{array}\right), \nonumber \\
{\rm Cor}(D^+) &=\left(\begin{array}{ccccc}
1 . & -0.00918393 & -0.0111696 & -0.011663 & -0.0114986 \\
-0.00918393 & 1 . & 0.959852 & 0.877876 & 0.784972 \\
-0.0111696 & 0.959852 & 1 . & 0.974619 & 0.917145 \\
-0.011663 & 0.877876 & 0.974619 & 1 . & 0.982034 \\
-0.0114986 & 0.784972 & 0.917145 & 0.982034 & 1 .
\end{array}\right),\\
{\rm Cor}(D^+_s) &=\left(\begin{array}{ccccc}
1 . & -0.0948363 & -0.0854423 & -0.0692735 & -0.0528099 \\
-0.0948363 & 1 . & 0.961103 & 0.877715 & 0.779303 \\
-0.0854423 & 0.961103 & 1 . & 0.973193 & 0.910118 \\
-0.0692735 & 0.877715 & 0.973193 & 1 . & 0.979699 \\
-0.0528099 & 0.779303 & 0.910118 & 0.979699 & 1 .
\end{array}\right). \nonumber 
\end{align}
The second set includes the decay width, the fist order electron energy moment, the second, third and the fourth electron energy center moments. Their correlation matrix for the $D^0$, $D^+$ and $D_s^+$ decays are given by
\begin{align}\label{eq:corcenmoments}
{\rm Cor}(D^0) &=\left(\begin{array}{ccccc}
1 . & -0.0764818 & 0.0281587 & 0.0330765 & 0.0102788 \\
-0.0764818 & 1 . & -0.088841 & -0.444378 & -0.0618813 \\
0.0281587 & -0.088841 & 1 . & -0.00420729 & 0.817582 \\
0.0330765 & -0.444378 & -0.00420729 & 1 . & -0.0351105 \\
0.0102788 & -0.0618813 & 0.817582 & -0.0351105 & 1 .
\end{array}\right), \nonumber\\
{\rm Cor}(D^+) &=\left(\begin{array}{ccccc}
1 . & -0.00918393 & -0.00799742 & 0.00816853 & -0.0107153 \\
-0.00918393 & 1 . & -0.0423819 & -0.456701 & -0.0390424 \\
-0.00799742 & -0.0423819 & 1 . & -0.0538756 & 0.809543 \\
0.00816853 & -0.456701 & -0.0538756 & 1 . & -0.0924318 \\
-0.0107153 & -0.0390424 & 0.809543 & -0.0924318 & 1 .
\end{array}\right), \\
{\rm Cor}(D^+_s) &=\left(\begin{array}{ccccc}
1 . & -0.0948363 & 0.0254998 & 0.0800786 & 0.00809594 \\
-0.0948363 & 1 . & -0.047314 & -0.40486 & -0.0191632 \\
0.0254998 & -0.047314 & 1 . & 0.0763575 & 0.800281 \\
0.0800786 & -0.40486 & 0.0763575 & 1 . & 0.172337 \\
0.00809594 & -0.0191632 & 0.800281 & 0.172337 & 1 .
\end{array}\right) . \nonumber
\end{align}

\end{appendix}

\bibliography{Charminclusivedecay.bib}



\end{document}